\documentclass[subeqn,11pt, a4paper]{amsart}
\usepackage[margin=35mm]{geometry}

\usepackage{amssymb,amsmath,amsfonts,amsthm, amscd, mathrsfs, helvet, mathtools}
\usepackage{bm, stmaryrd}
\usepackage{graphicx, verbatim}
\usepackage[mathcal]{euscript}
\usepackage{color}
\usepackage[all]{xy}
\usepackage{relsize}

\usepackage{tikz}
\usetikzlibrary{arrows,matrix,calc}
\usetikzlibrary{decorations.markings,shapes.geometric,shapes.misc}
\usetikzlibrary{decorations.pathreplacing,positioning}  

\tikzset{cross/.style={cross out, draw=black, minimum size=2*(#1-\pgflinewidth), inner sep=0pt, outer sep=0pt}, cross/.default={1pt}}

\allowdisplaybreaks[1]

\usepackage{pict2e}



\definecolor{myPurple}{rgb}{0.5,0.1,0.6}
\definecolor{myOrange}{rgb}{1.0,0.5,0.0}
\definecolor{myRed}{rgb}{1.0,0.0,0.0}
\definecolor{myGreen}{rgb}{0.0,0.5,0.0}
\definecolor{LatexBlue}{rgb}{0.211765,0.227451,0.666667}
\definecolor{myBlue}{rgb}{0.0,0.0,1.0}
\definecolor{myBlack}{rgb}{0.0,0.0,0.0}
\definecolor{myGray}{rgb}{0.3,0.3,0.3}

\theoremstyle{plain}

\newtheorem*{theorem*}{Theorem}

\newtheorem*{proposition*}{Proposition}

\theoremstyle{remark}

\newtheorem{remarkx}{Remark}[section]

\newenvironment{remark}
  {\pushQED{\qed}\remarkx}
  {\popQED\endremarkx}

\newcommand{\tensor}[1]{{\mathfrak{#1}}}

\DeclareMathOperator{\res}{res}

\def\g{\mathfrak{g}}

\def\id{\textup{id}}

\def\cent{\mathsf{K}}
\def\cocent{\mathsf{D}}

\def\tg{\widetilde{\mathfrak{g}}}

\def\L{\mathcal{L}}

\def\ha{\mbox{\small $\frac{1}{2}$}}

\newcommand{\lau}[1]{(\kern-.2em( #1 )\kern-.2em)}%{(( #1 ))}

\def\CC{\mathbb{C}}
\def\CP{\mathbb{C}P^1}

\def\RR{\mathbb{R}}

\def\ZZ{\mathbb{Z}}
\def\H{\mathcal{H}}

\def\P{\mathcal{P}}
\def\R{\mathcal{R}}

\def\ii{{\rm i}}

\def\1{\tensor{1}}
\def\2{\tensor{2}}
\def\3{\tensor{3}}
\def\4{\tensor{4}}

\numberwithin{equation}{section}

\begin{document}

\title[Holomorphic Chern-Simons theory and affine Gaudin models]{Holomorphic Chern-Simons theory\\[1mm]
and affine Gaudin models}

\author{Beno\^{\i}t Vicedo}
\address{Department of Mathematics, University of York, York YO10 5DD, U.K.} \email{benoit.vicedo@gmail.com}

\begin{abstract}
We relate two formalisms recently proposed for describing classical integrable field theories. The first \cite{Costello:2019tri} is based on the action of four-dimensional holomorphic Chern-Simons theory introduced and studied by Costello, Witten and Yamazaki. The second \cite{Vicedo:2017cge} makes use of classical generalised Gaudin models associated with untwisted affine Kac-Moody algebras.
\end{abstract}

\maketitle

%\setcounter{tocdepth}{1}
%\tableofcontents

\input{epsf}

\section{Introduction and summary}

It was shown by Costello in \cite{Costello:2013zra, Costello:2013sla}, and further developed recently in \cite{Witten:2016spx, Costello:2017dso,Costello:2018gyb} by Costello, Witten and Yamazaki, that various integrable lattice models can be understood as originating from a four-dimensional variant of Chern-Simons theory on the product $M \coloneqq \Sigma \times C$ of a real two-dimensional manifold $\Sigma$ and a Riemann surface $C$ equipped with a \emph{non-vanishing} meromorphic 1-form $\omega$. It was also recently shown in \cite{Bittleston:2019gkq} that integrable lattice models with boundaries can be accounted for by putting the gauge theory on an orbifold $(\Sigma \times \CC)/\ZZ_2$.

Very recently in \cite{Costello:2019tri}, Costello and Yamazaki extended this approach to describe also integrable field theories on $\Sigma$, with spectral plane $C$, by starting from the \emph{same} variant of Chern-Simons theory on $\Sigma \times C$ as in \cite{Costello:2013zra, Costello:2013sla, Witten:2016spx, Costello:2017dso,Costello:2018gyb}.

\medskip

The purpose of this note is to show that the framework of \cite{Costello:2019tri} is intimately related to the description of classical integrable field theories that we proposed in \cite{Vicedo:2017cge}, which is based on Gaudin models associated with untwisted affine Kac-Moody algebras. In order to explain this connection in more detail, we first recall how the gauge theory on $\Sigma \times C$ is defined more explicitly.

\medskip

For concreteness, we shall let $\Sigma = \RR \times S^1$ with global coordinates $(\tau, \sigma)$ and let $C = \CP$ be the Riemann sphere with holomorphic coordinate $z$ on $\CC = \CP \setminus \{ \infty \}$. We also fix a choice of meromorphic differential $\omega$ on $\CP$ which can be expressed in coordinates as
\begin{equation*}
\omega = \varphi(z) dz,
\end{equation*}
where $\varphi$ is a meromorphic function on $\CP$. As noted in \cite{Costello:2019tri}, in order to be able to describe a broad family of classical integrable field theories it is crucial in the present context to allow $\omega$ to have zeroes.

Let $\g$ be a semisimple Lie algebra over $\CC$ and $\langle \cdot, \cdot \rangle : \g \times \g \to \CC$ be a non-degenerate invariant symmetric bilinear form on $\g$. We extend it to a symmetric bilinear pairing $\langle \cdot, \cdot \rangle : \g \otimes \Omega^p(M) \times \g \otimes \Omega^q(M) \to \Omega^{p+q}(M)$.

The bulk action functional of \emph{holomorphic Chern-Simons theory} introduced and studied in \cite{Costello:2013zra, Costello:2013sla, Witten:2016spx, Costello:2017dso,Costello:2018gyb,Costello:2019tri}, for a $\g$-valued $1$-form $A \in \g \otimes \Omega^1(M)$, reads
\begin{equation} \label{holo CS action}
S_{\rm bulk}[A] = \frac{\ii}{4\pi} \int_{\Sigma \times \CP} \omega \wedge CS(A), \qquad CS(A) \coloneqq \langle A, dA + \mbox{\small $\frac{2}{3}$} A \wedge A \rangle.
\end{equation}
The normalisation factor in front of the action is chosen to match the conventions of \cite{Vicedo:2017cge}. It is interesting to note that it coincides, up to an integer factor, with the normalisation of the action (in the case when $\omega = dz$) motivated from the extension of the standard Chern-Simons action to loop groups \cite{Witten:2016spx}.

The action \eqref{holo CS action} is trivially
invariant under the transformation $A \mapsto A + \varpi$ for any $\varpi \in \Omega^{1,0}(\CP)$. We can use this freedom to eliminate the $(1,0)$-component of $A$ along $\CP$ and thereby fix it to be of the form $A = A_\tau d\tau + A_\sigma d\sigma + A_{\bar z} d\bar z$. The action is then invariant under gauge transformations of these remaining three components.

\medskip

The equation of motion $\omega \wedge F = 0$, derived by extremising \eqref{holo CS action}, expresses the fact that $A$ is a flat connection on $\Sigma$ which varies holomorphically on $\CP$. This strongly suggests that $A_\tau d\tau + A_\sigma d\sigma$ can be interpreted as the Lax connection of some classical integrable field theory.
Indeed, the proposal of \cite{Costello:2019tri} is to describe various integrable field theories as arising from the introduction of specific surface defects along $\Sigma$ in the holomorphic Chern-Simons theory on $M = \Sigma \times \CP$.

However, in order to completely characterise the integrable structure of a classical integrable field theory, it is necessary to move to the Hamiltonian framework and to identify the Poisson bracket of $A_\sigma$ with itself. There then exists sufficient conditions on the form of this Poisson bracket \cite{Maillet1, Maillet2} ensuring that the integrals of motion constructed from $A_\sigma$ are in involution.

\medskip

In \S\ref{sec: Ham CS theory} we perform a Hamiltonian analysis of the holomorphic Chern-Simons theory of \cite{Costello:2013zra, Costello:2013sla, Witten:2016spx, Costello:2017dso,Costello:2018gyb, Costello:2019tri}, with $\omega$ a generic meromorphic differential on $\CP$. There are first class constraints associated with the gauge invariance of this theory and second class constraints coming from the fact that the Lagrangian $CS(A)$ is linear in the time derivative of $A$. We impose natural gauge fixing conditions and determine the corresponding Dirac bracket $\{ \cdot, \cdot \}^\star$ on the reduced phase space. The latter is parametrised by the $\g$-valued field $A_\sigma$ which, having fixed the gauge, is now \emph{meromorphic} and such that:
\begin{subequations} \label{Gaudin intro}
\begin{equation} \label{local condition}
\textup{the combination $\varphi A_\sigma$ has the same pole structure as $\varphi$.}
\end{equation}

We find that the Dirac bracket on the reduced phase space
takes the form
\begin{align} \label{rs algebra intro}
\{ A_{\sigma \1}(z,\sigma), A_{\sigma \2}(z', \sigma') \}^\star 
&= \big[ \R_{\1\2}(z, z'), A_{\sigma \1}(z, \sigma) \big] \delta_{\sigma\sigma'} - \big[ \R_{\2\1}(z', z), A_{\sigma \2}(z', \sigma) \big] \delta_{\sigma\sigma'} \notag\\
&\qquad\qquad\qquad\qquad\qquad - \big( \R_{\1\2}(z, z') + \R_{\2\1}(z', z) \big) \delta'_{\sigma\sigma'},
\end{align}
where the $\R$-matrix is given explicitly by
\begin{equation} \label{R matrix intro}
\R_{\1\2}(z, z') \coloneqq 2\pi \frac{C_{\1\2}}{z' - z} \varphi(z')^{-1}.
\end{equation}
The factor of $2 \pi$ in \eqref{R matrix intro} is also there to match the conventions of \cite{Vicedo:2017cge}.

In other words, the first result of this note is that the spatial component $A_\sigma$ of the Chern-Simons 1-form $A$ can be interpreted as the Lax matrix of a \emph{non-ultralocal} classical integrable field theory with \emph{twist function} $\varphi$.

Furthermore, by adding to the bulk Hamiltonian associated with the action \eqref{holo CS action} a suitable boundary term, fixed by the requirement that the total Hamiltonian has well-defined functional derivatives \cite{Regge:1974zd}, we find that the Hamiltonian on the reduced phase space is
\begin{equation} \label{Ham intro}
H = -\frac{1}{2} \sum_{x \in \bm \zeta} \epsilon_x \int_{S^1} d\sigma \, \res_x \langle A_\sigma, A_\sigma \rangle \omega,
\end{equation}
\end{subequations}
where $\bm \zeta$ is the set of zeroes of $\omega$ and $\{ \epsilon_x \}_{x \in \bm \zeta}$ is a set of complex numbers entering through the choice of gauge fixing conditions imposed.

\medskip

Let $\tg$ be the untwisted affine Kac-Moody algebra corresponding to $\g$. We showed in \cite{Vicedo:2017cge} (see also \cite{Delduc:2019bcl, Lacroix:2019xeh}) that classical integrable field theories with the properties \eqref{rs algebra intro} \--- \eqref{Ham intro} can be understood as realisations of various generalisations of the Gaudin model associated with $\tg$. Since this result is quite technical, but essential to the discussion, we will recall its main features in \S\ref{sec: affine Gaudin}.

The result of this note therefore establishes that the general formalisms of \cite{Costello:2019tri} and \cite{Vicedo:2017cge} provide equivalent descriptions of classical integrable field theories in the Lagrangian and Hamiltonian formulations, respectively.

More precisely, the action functional \eqref{holo CS action} of \cite{Costello:2019tri} can be used to describe those classical integrable field theories which:
\begin{itemize}
  \item[$(i)$] can be realised as a non-cyclotomic affine Gaudin model in the sense of \cite{Vicedo:2017cge},
  \item[$(ii)$] satisfy the additional technical condition \eqref{local condition}.
\end{itemize}
We do not discuss here the case of cyclotomic affine Gaudin model, let alone dihedral ones, in the terminology of \cite{Vicedo:2017cge}. See \S\ref{sec: dihedral} for a discussion of this point.

It is, however, interesting to note that the condition \eqref{local condition} is known not to hold for certain classical integrable field theories which nevertheless do admit an affine Gaudin model description. This is, for instance, the case for affine Toda field theories which can be described as cyclotomic (in fact dihedral) affine Gaudin models \cite{Vicedo:2017cge}. The generalisation of the present work to the cyclotomic case could therefore provide an explanation as to why these theories, including sine-Gordon theory, do not admit straightforward interpretations in terms of holomorphic Chern-Simons theory \cite{Costello:2019tri}.

\medskip

We end with some comments and discussion of possible future work in \S\ref{sec: discussion}.

\subsubsection*{Acknowledgements}

I would like to thank Roland Bittleston, Sylvain Lacroix, Carlo Meneghelli and Masahito Yamazaki for useful discussions.

\section{Hamiltonian analysis of holomorphic Chern-Simons theory} \label{sec: Ham CS theory}

\subsection{Bulk action}

In order to move to the Hamiltonian framework we begin by isolating the global time coordinate on the cylinder by writing $A = A_\tau d\tau + \hat A$ with $\hat A \coloneqq A_\sigma d\sigma + A_{\bar z} d \bar z$, and for any $\eta \in \g \otimes \Omega^p(M)$ we let $d\eta = d\tau \wedge \partial_\tau \eta + \hat d\eta$, with $\hat d\eta \coloneqq d\sigma \wedge \partial_\sigma \eta + dz \wedge \partial_z \eta + d\bar z \wedge \partial_{\bar z} \eta$.

We have
\begin{equation*}
CS(A) = - d\tau \wedge \big( \langle \hat A, \partial_\tau \hat A \rangle - 2 \langle A_\tau, \hat F \rangle \big) + \hat d \langle A_\tau d\tau, \hat A \rangle + \langle \hat A, \hat d \hat A \rangle,
\end{equation*}
where $\hat F \coloneqq \hat d \hat A + \hat A \wedge \hat A$. The last term in $CS(A)$ can be ignored since it will drop out when taking the wedge product with $\omega$. The bulk action functional \eqref{holo CS action} can then be rewritten as
\begin{equation} \label{action v0}
S_{\rm bulk}[A] = \frac{\ii}{4\pi} \int_{\RR \times S^1 \times \CP} d\tau \wedge \omega \wedge \big( \langle \hat A, \partial_\tau \hat A \rangle - 2 \langle A_\tau, \hat F \rangle \big)
\end{equation}
where we ignored a `boundary term'. Indeed, even though $S^1 \times \CP$ has no boundary \emph{per se}, using Stokes's theorem we generate a term of the form $d \tau \wedge \hat d \omega \wedge \langle A_\tau, \hat A \rangle$ in the integrand. Explicitly, we have
\begin{equation*}
\int_{S^1 \times \CP} \omega \wedge \hat d \langle A_\tau, \hat A \rangle = - \int_{S^1 \times \CP} \hat d\big(\omega \wedge \langle A_\tau, \hat A \rangle\big) + \int_{S^1 \times \CP} \hat d \omega \wedge \langle A_\tau, \hat A \rangle,
\end{equation*}
with the first term vanishing because $\partial (S^1 \times \CP) = \emptyset$. But $\hat d \omega$ is a distribution on $\CP$ with support at the poles of $\omega$, so the integral in the second term above localises at these poles. We shall therefore refer to such terms as `boundary terms'.

\begin{remark} \label{rem: boundary terms}
One could equally describe these `boundary terms' as \emph{actual} boundary terms. Let $D_r$ be the union of small discs of radius $r > 0$ around each of the poles of $\omega$, and take the integral in \eqref{action v0} to be over $M_r \coloneqq S^1 \times \CP \setminus D_r$ instead. Then
\begin{equation*}
\int_{M_r} \omega \wedge \hat d \langle A_\tau, \hat A \rangle = - \int_{M_r} \hat d\big(\omega \wedge \langle A_\tau, \hat A \rangle\big) + \int_{M_r} \hat d \omega \wedge \langle A_\tau, \hat A \rangle,
\end{equation*}
where now the second term on the right hand side vanishes because $\hat d \omega$ has support inside $D_r$. On the other hand, the first term now gives a boundary integral which in the limit when $r \to 0$ coincides with the `boundary term' identified above. More generally, in order to allow other singularities in the fields $A_\tau$ or $\hat A$, as we will do when imposing a gauge fixing condition on $A_\tau$ later in \S\ref{sec: fix Lagr}, we should also include in $D_r$ small discs of radius $r$ around these additional points.
\end{remark}

Writing $\hat A$ in terms of its components and working up to `boundary terms' in the above sense, we can express the action \eqref{action v0} more explicitly as
\begin{subequations} \label{S bulk}
\begin{align}
S_{\rm bulk}[A] &= \int_{\RR \times S^1 \times \CP} d\tau \wedge d\sigma \wedge dz \wedge d\bar z \, \L_{\rm bulk}(A),
\end{align}
where the bulk Lagrangian is given by
\begin{align} \label{bulk Lagrangian}
\L_{\rm bulk}(A) &\coloneqq \frac{\ii \varphi}{4\pi} \langle A_{\bar z}, \partial_\tau A_\sigma \rangle - \frac{\ii \varphi}{4\pi} \langle A_\sigma, \partial_\tau A_{\bar z} \rangle \notag\\
&\qquad\qquad\qquad - \frac{\ii}{2\pi} \langle A_\tau, \partial_{\bar z} (\varphi A_\sigma) - \varphi \partial_\sigma A_{\bar z} - [\varphi A_\sigma, A_{\bar z}] \rangle.
\end{align}
\end{subequations}

\subsection{Phase space} \label{sec: phase space}

The conjugate momentum of the three $\g$-valued fields $A_\tau$, $A_\sigma$ and $A_{\bar z}$ are given respectively by the $\g$-valued fields
\begin{equation*}
\Pi_\tau \coloneqq \frac{\delta \mathcal L(A)}{\delta (\partial_\tau A_\tau)} = 0, \qquad
\Pi_\sigma \coloneqq \frac{\delta \mathcal L(A)}{\delta (\partial_\tau A_\sigma)} = \frac{\ii \varphi}{4 \pi} A_{\bar z}, \qquad
\Pi_{\bar z} \coloneqq \frac{\delta \mathcal L(A)}{\delta (\partial_\tau A_{\bar z})} = - \frac{\ii \varphi}{4 \pi} A_\sigma.
\end{equation*}

The initial phase space is parametrised by three pairs of $\g$-valued conjugate fields $A_i, \Pi_i \in C^\infty(S^1 \times \CP, \g)$ for $i \in \{ \tau, \sigma, \bar z \}$, whose canonical Poisson brackets can be expressed using standard tensorial index notation as
\begin{equation} \label{original PB}
\{ A_{i \1}(\sigma, z), \Pi_{i \2}(\sigma', z') \} = 2 \pi C_{\1\2} \delta_{\sigma\sigma'} \delta_{zz'},
\end{equation}
where $\delta_{\sigma\sigma'} \coloneqq \frac{1}{2\pi} \sum_{n \in \ZZ} e^{\ii n (\sigma - \sigma')}$ is the Dirac comb, \emph{i.e.} the Dirac $\delta$-distribution on $S^1$, and $\delta_{zz'}$ is the Dirac $\delta$-distribution on $\CP$ with the properties that
\begin{equation*}
\int_{S^1} d\sigma \, f(\sigma, z) \delta_{\sigma\sigma'} = f(\sigma', z), \qquad
\int_{\CP} dz \wedge d\bar z \, f(\sigma, z) \delta_{zz'} = f(\sigma, z')
\end{equation*}
for any $f \in C^\infty(S^1 \times \CP)$. Also, $C$ denotes the split Casimir of $\g$.

There are three primary constraints
\begin{equation} \label{primary constraints}
\Pi_\tau \approx 0, \qquad
\mathcal C_\sigma \coloneqq A_{\bar z} - \frac{4 \pi}{\ii \varphi} \Pi_\sigma \approx 0, \qquad
\mathcal C_{\bar z} \coloneqq \Pi_{\bar z} + \frac{\ii \varphi}{4 \pi} A_\sigma \approx 0.
\end{equation}
The last two constraints are second class and their Poisson bracket
\begin{equation*}
\{ \mathcal C_{\sigma \1}(\sigma, z), \mathcal C_{\bar z \2}(\sigma', z') \} = 4 \pi C_{\1\2} \delta_{\sigma\sigma'} \delta_{zz'}
\end{equation*}
is invertible. We can therefore set them to zero strongly, which we shall do, provided that we work with the corresponding Dirac brackets, given by
\begin{subequations} \label{A Pi bar z PB}
\begin{align}
\label{A Pi bar z PB a} \{ A_{\tau \1}(\sigma, z), \Pi_{\tau \2}(\sigma', z') \}^\ast &= 2 \pi C_{\1\2} \delta_{\sigma\sigma'} \delta_{zz'}, \\
\label{A Pi bar z PB b} \{ A_{\bar z \1}(\sigma, z), \Pi_{\bar z \2}(\sigma', z') \}^\ast &= \pi C_{\1\2} \delta_{\sigma\sigma'} \delta_{zz'}.
\end{align}
\end{subequations}

Let $\mathcal P$ denote the resulting phase space, parametrised by the fields $A_\tau$, $\Pi_\tau$, $A_{\bar z}$ and $\Pi_{\bar z}$ satisfying the Dirac brackets \eqref{A Pi bar z PB}. We shall refer to the latter just as a Poisson bracket from now on, but still keep denoting it as $\{ \cdot, \cdot \}^\ast$ to distinguish it from the original Poisson bracket \eqref{original PB} since \eqref{A Pi bar z PB b} is now different.

Note that we have thus far fixed the last two of the primary constraints in \eqref{primary constraints}, so there remains the primary constraint $\Pi_\tau \approx 0$.

\subsection{Differentiable functionals} \label{sec: diff func}

Given any pair of functionals $\mathscr F, \mathscr G : \P \to \CC$, it follows from \eqref{A Pi bar z PB} that their Poisson bracket reads
\begin{align} \label{PB of functionals}
\{ \mathscr F, \mathscr G \}^\ast &= 2 \pi \bigg\langle\!\!\!\bigg\langle \frac{\delta \mathscr F}{\delta A_\tau}, \frac{\delta \mathscr G}{\delta \Pi_\tau} \bigg\rangle\!\!\!\bigg\rangle - 2 \pi \bigg\langle\!\!\!\bigg\langle \frac{\delta \mathscr F}{\delta \Pi_\tau}, \frac{\delta \mathscr G}{\delta A_\tau} \bigg\rangle\!\!\!\bigg\rangle \notag\\
&\qquad\qquad\qquad + \pi \bigg\langle\!\!\!\bigg\langle \frac{\delta \mathscr F}{\delta A_{\bar z}}, \frac{\delta \mathscr G}{\delta \Pi_{\bar z}} \bigg\rangle\!\!\!\bigg\rangle - \pi \bigg\langle\!\!\!\bigg\langle \frac{\delta \mathscr F}{\delta \Pi_{\bar z}}, \frac{\delta \mathscr G}{\delta A_{\bar z}} \bigg\rangle\!\!\!\bigg\rangle,
\end{align}
where we have introduced the notation
\begin{equation} \label{double angle brackets}
\langle\!\langle X, Y \rangle\!\rangle \coloneqq \int_{S^1 \times \CP} d\sigma \wedge dz \wedge d\bar z \, \langle X, Y \rangle
\end{equation}
for any $\g$-valued distributions $X, Y$ on $S^1 \times \CP$ for which this integral makes sense.

However, problems could arise if the variational derivatives of $\mathscr F$ and $\mathscr G$ involve distributions with overlapping supports. The right hand side of \eqref{PB of functionals} would then be the integral of a product of such distributions, which is typically ill-defined.

\medskip

In light of Remark \ref{rem: boundary terms}, such distributions can be interpreted as `boundary terms'. The treatment of boundary terms in the Hamiltonian framework was understood in the seminal work of Regge and Teitelboim \cite{Regge:1974zd} in the context of general relativity; see also \cite{Brown:1986nw, Brown:1986ed}. The application of these ideas to ordinary Chern-Simons theory, directly relevant to the present discussion, was considered in \cite{Banados:1994tn, Banados:1998gg}; see also \cite{Banados:2016zim}.

We shall say, in the spirit of \cite{Regge:1974zd}, that a functional $\mathscr F : \P \to \CC$ is \emph{differentiable} if its variational derivatives do not involve `boundary terms', \emph{i.e.} if we can write
\begin{equation*}
\delta \mathscr F = \bigg\langle\!\!\!\bigg\langle \frac{\delta \mathscr F}{\delta A_\tau}, \delta A_\tau \!  \bigg\rangle\!\!\!\bigg\rangle + \bigg\langle\!\!\!\bigg\langle \frac{\delta \mathscr F}{\delta A_{\bar z}}, \delta A_{\bar z} \!  \bigg\rangle\!\!\!\bigg\rangle + \bigg\langle\!\!\!\bigg\langle \frac{\delta \mathscr F}{\delta \Pi_\tau}, \delta \Pi_\tau \!  \bigg\rangle\!\!\!\bigg\rangle + \bigg\langle\!\!\!\bigg\langle \frac{\delta \mathscr F}{\delta \Pi_{\bar z}}, \delta \Pi_{\bar z} \!  \bigg\rangle\!\!\!\bigg\rangle
\end{equation*}
but where the variational derivatives $\delta \mathscr F/ \delta A_i$ and $\delta \mathscr F/ \delta \Pi_i$ are \emph{smooth} functions for $i \in \{ \tau, \bar z \}$, though possibly with singularities at finitely many points.

The resolution of the problem alluded to above is that the Poisson bracket $\{ \mathscr F, \mathscr G \}^\ast$ is only defined between differentiable functionals $\mathscr F$ and $\mathscr G$. If a functional $\mathscr F$ is not differentiable then one should find a suitable boundary term to add to it, so as to cancel off any unwanted boundary terms in its variation $\delta \mathscr F$. This will ensure that it has well-defined Poisson brackets with any other differentiable functional.

\subsection{Bulk Hamiltonian} \label{sec: bulk Ham}

The bulk Hamiltonian density is given by the Legendre transform of the bulk Lagrangian \eqref{bulk Lagrangian}, namely
\begin{equation*}
\H_{\rm bulk}(A) \coloneqq \langle \Pi_\tau, \partial_\tau A_\tau \rangle + \langle \Pi_\sigma, \partial_\tau A_\sigma \rangle + \langle \Pi_{\bar z}, \partial_\tau A_{\bar z} \rangle
- \mathcal L_{\rm bulk}(A) = \langle A_\tau, \gamma \rangle.
\end{equation*}
Here we have introduced the $\g$-valued field
\begin{equation*}
\gamma \coloneqq - 2 \partial_{\bar z} \Pi_{\bar z} - \mbox{\small $\frac{\ii}{2 \pi}$} \varphi \, \partial_\sigma A_{\bar z} + 2 [\Pi_{\bar z}, A_{\bar z}].
\end{equation*}
Therefore, the bulk Hamiltonian of holomorphic Chern-Simons theory simply reads
\begin{equation} \label{Hamiltonian v0}
H_{\rm bulk} \coloneqq \langle\!\langle A_\tau, \gamma \rangle\!\rangle.
\end{equation}
We will come back to the issue of the differentiability of this functional in \S\ref{sec: reduced dynamics} after fixing the value of $A_\tau$ in \S\ref{sec: fix Lagr}.

\subsection{Gauge invariance} \label{sec: gauge inv}

There is just one constraint on the phase space $\P$, namely $\Pi_\tau \approx 0$, which we should ensure is preserved under time evolution. We have
\begin{equation*}
\{ H_{\rm bulk}, \Pi_\tau \}^\ast = \gamma.
\end{equation*}
This is, however, not quite a pure constraint since it already contains the necessary `boundary terms' to ensure that $\langle\! \langle \varepsilon, \gamma \rangle\!\rangle$, for all $\varepsilon \in C^\infty(S^1 \times \CP, \g)$, is a differentiable functional in the sense of \S\ref{sec: diff func}, cf. the computation in \S\ref{sec: reduced dynamics} below.

On the other hand, $\gamma$ is the correct `improved' generator of gauge transformations. Indeed, by using \eqref{A Pi bar z PB b} we obtain
\begin{subequations}
\begin{align}
\label{PB gamma Asig} \mbox{\small $\frac{1}{2\pi}$} \{ \gamma_\1(\sigma, z), A_{\sigma \2}(\sigma', z') \}^\ast &= [C_{\1\2}, A_{\sigma \2}(\sigma, z)] \delta_{\sigma\sigma'} \delta_{zz'} + C_{\1\2} \delta'_{\sigma\sigma'} \delta_{zz'},\\
\label{PB gamma Azb} \mbox{\small $\frac{1}{2\pi}$} \{ \gamma_\1(\sigma, z), A_{\bar z \2}(\sigma', z') \}^\ast &= [C_{\1\2}, A_{\bar z \2}(\sigma, z)] \delta_{\sigma\sigma'} \delta_{zz'} + C_{\1\2} \delta_{\sigma\sigma'} \partial_{\bar z} \delta_{zz'}.
\end{align}
\end{subequations}
It follows that the expression $\mbox{\small $\frac{1}{2\pi}$} \langle\!\langle \varepsilon, \gamma \rangle\!\rangle$, for every $\varepsilon \in C^\infty(S^1 \times \CP, \g)$, generates a gauge transformation of holomorphic Chern-Simons theory since
\begin{align*}
\mbox{\small $\frac{1}{2\pi}$} \{ \langle\!\langle \varepsilon, \gamma \rangle\!\rangle, A_\sigma(\sigma, z) \}^\ast &= [\varepsilon(\sigma, z), A_\sigma(\sigma, z)] - \partial_\sigma \varepsilon(\sigma, z),\\
\mbox{\small $\frac{1}{2\pi}$} \{ \langle\!\langle \varepsilon, \gamma \rangle\!\rangle, A_{\bar z}(\sigma, z) \}^\ast &= [ \varepsilon(\sigma, z), A_{\bar z}(\sigma, z)] - \partial_{\bar z} \varepsilon(\sigma, z).
\end{align*}
In particular, the bulk Hamiltonian \eqref{Hamiltonian v0} is thus a pure gauge transformation with the field $A_\tau$ playing the role of the gauge parameter.

Moreover, the Poisson bracket of $\gamma$ with itself reads
\begin{equation} \label{gamma not FC}
\{ \gamma_\1(\sigma, z), \gamma_\2(\sigma', z') \}^\ast = 2 \pi [ C_{\1\2}, \gamma_\2(\sigma, z) ] \delta_{\sigma\sigma'} \delta_{zz'} + \ii (\partial_{\bar z} \varphi(z)) C_{\1\2} \delta'_{\sigma\sigma'} \delta_{zz'},
\end{equation}
from which it follows that, for any $\varepsilon, \tilde \varepsilon \in C^\infty(S^1 \times \CP, \g)$, we have
\begin{equation} \label{gamma not FC epsilon}
\{ \langle\!\langle \varepsilon, \gamma \rangle\!\rangle, \langle\!\langle \tilde\varepsilon, \gamma \rangle\!\rangle \}^\ast = - 2 \pi \langle\!\langle [\varepsilon, \tilde \varepsilon], \gamma \rangle\!\rangle + \ii \langle\!\langle (\partial_{\bar z} \varphi) \varepsilon, \partial_\sigma \tilde\varepsilon \rangle\!\rangle.
\end{equation}
The second term on the right hand side is a `boundary term' localised at the poles of the differential $\omega$, cf. the analogous central extension in the Poisson algebra of the `improved' constraints in ordinary Chern-Simons theory \cite{Banados:1994tn, Banados:1998gg}.

Let $C^\infty(S^1 \times \CP, \g)_\omega$ denote the subspace of $C^\infty(S^1 \times \CP, \g)$ consisting of those functions which vanish at the poles of $\omega$ (and whose multiplicities at these zeroes is given by the orders of the corresponding poles of $\omega$).
The central extension term in \eqref{gamma not FC epsilon} is then absent if either $\varepsilon$ or $\tilde \varepsilon$ belongs to $C^\infty(S^1 \times \CP, \g)_\omega$. In particular, we see that the true bulk constraint can be described as the smearing $\langle\!\langle \varepsilon, \gamma \rangle\!\rangle \approx 0$ with all possible $\varepsilon \in C^\infty(S^1 \times \CP, \g)_\omega$. This is then first class by \eqref{gamma not FC epsilon}. By an abuse of language we will still refer to this constraint as $\gamma \approx 0$.

Using \eqref{gamma not FC} we find that
\begin{equation*}
\{ H_{\rm bulk}, \gamma \}^\ast \approx - \ii (\partial_{\bar z} \varphi) \partial_\sigma A_\tau.
\end{equation*}
Thus, for each $\varepsilon \in C^\infty(S^1 \times \CP, \g)_\omega$ we have $\{ H_{\rm bulk}, \langle\!\langle \varepsilon, \gamma \rangle\!\rangle \}^\ast \approx 0$, and hence there are no tertiary constraints.

\subsection{Gauge fixing} \label{sec: gauge fixing}

We would like to fix the gauge invariance associated with the constraint $\gamma \approx 0$ identified in \S\ref{sec: gauge inv}. Concretely, letting $\bm z$ denote the set of poles of $\varphi$ we will fix the constraint $\gamma(\sigma, z) \approx 0$ for $z \not\in \bm z$, which is clearly first class by \eqref{gamma not FC}.

\medskip

We shall do this by imposing the gauge fixing condition
\begin{equation} \label{holomorphic gauge}
A_{\bar z} \approx 0.
\end{equation}
It follows from \eqref{PB gamma Azb} that
\begin{equation} \label{PB Azb F}
\{ A_{\bar z \1}(\sigma, z), \gamma_\2(\sigma', z') \}^\ast = \{ \gamma_\1(\sigma, z), A_{\bar z \2}(\sigma', z') \}^\ast \approx - 2 \pi \partial_{\bar z'} ( C_{\1\2} \delta_{\sigma\sigma'} \delta_{zz'} ).
\end{equation}
We can thus impose the constraint $\gamma \approx 0$ together with the gauge fixing condition \eqref{holomorphic gauge} strongly, provided that we work with the appropriate new Dirac bracket $\{ \cdot, \cdot \}^\star$. To define it, we note that \eqref{PB Azb F} is invertible since
\begin{equation*}
\bigg\langle\!\!\!\bigg\langle \!\! - 2 \pi \partial_{\bar z'} ( C_{\1\2} \delta_{\sigma\sigma'} \delta_{zz'} ), - \frac{1}{(2 \pi)^2 \ii} \frac{C_{\2\3} \delta_{\sigma'\sigma''}}{z' - z''} \bigg\rangle\!\!\!\bigg\rangle_{(\sigma', z') \2} = C_{\1\3} \delta_{\sigma\sigma''} \delta_{zz''}.
\end{equation*}
The subscript `$(\sigma', z') \2$' on $\langle\!\langle \cdot, \cdot \rangle\!\rangle$, as defined in \eqref{double angle brackets}, is used here to indicate that the integration is taken over $d\sigma' \wedge dz' \wedge d\bar z'$ and the bilinear form $\langle \cdot, \cdot \rangle$ is applied to the second tensor factor.

The new Dirac bracket of any $\g$-valued observables $U$ and $V$ is then defined by
\begin{align*}
&\{ U_\1(\sigma, z), V_\2(\sigma', z') \}^\star \coloneqq \{ U_\1(\sigma, z), V_\2(\sigma', z') \}^\ast\\
&\qquad + 
\bigg\langle\!\!\!\bigg\langle \{ U_\1(\sigma, z), \gamma_\3(\sigma'', z'') \}^\ast,\\
&\qquad\qquad\qquad\qquad \bigg\langle\!\!\!\bigg\langle \frac{1}{(2 \pi)^2 \ii} \frac{C_{\3\4}\delta_{\sigma''\sigma'''}}{z'' - z'''}, \{ A_{\bar z\4}(\sigma''', z'''), V_\2(\sigma', z') \}^\ast \bigg\rangle\!\!\!\bigg\rangle_{(\sigma''', z''')\4} \bigg\rangle\!\!\!\bigg\rangle_{(\sigma'', z'')\3}\\
&\qquad + 
\bigg\langle\!\!\!\bigg\langle \{ U_\1(\sigma, z), A_{\bar z\3}(\sigma'', z'') \}^\ast,\\
&\qquad\qquad\qquad\qquad \bigg\langle\!\!\!\bigg\langle \frac{1}{(2 \pi)^2 \ii} \frac{C_{\3\4}\delta_{\sigma''\sigma'''}}{z'' - z'''}, \{ \gamma_\4(\sigma''', z'''), V_\2(\sigma', z') \}^\ast \bigg\rangle\!\!\!\bigg\rangle_{(\sigma''', z''')\4} \bigg\rangle\!\!\!\bigg\rangle_{(\sigma'', z'')\3}.
\end{align*}
In order to compute the Dirac bracket of the field $\Pi_{\bar z}$ with itself we note from \eqref{PB gamma Asig}, and using the last constraint in \eqref{primary constraints}, that
\begin{align*}
\{ \Pi_{\bar z \1}(\sigma, z), \gamma_\2(\sigma', z') \}^\ast &= \{ \gamma_\1(\sigma, z), \Pi_{\bar z \2}(\sigma', z') \}^\ast\\
&= 2\pi [C_{\1\2}, \Pi_{\bar z \2}(\sigma, z)] \delta_{\sigma\sigma'} \delta_{zz'} - \mbox{\small $\frac{\ii}{2}$} \varphi(z) C_{\1\2} \delta'_{\sigma\sigma'} \delta_{zz'}.
\end{align*}
Using the above definition for the Dirac bracket $\{ \cdot, \cdot \}^\star$ we obtain
\begin{align} \label{PB of Pi}
\frac{4 \pi}{\ii} \{ \Pi_{\bar z \1}(\sigma, z), \Pi_{\bar z\2}(\sigma', z') \}^\star 
&= \bigg[\frac{2 \pi C_{\1\2}}{z-z'}, \Pi_{\bar z \1}(\sigma, z) + \Pi_{\bar z \2}(\sigma', z') \bigg] \delta_{\sigma\sigma'} \notag\\
&\qquad\qquad\qquad\qquad + \ii \frac{2 \pi C_{\1\2}}{z-z'} \frac{\varphi(z) - \varphi(z')}{4 \pi} \delta'_{\sigma\sigma'}.
\end{align}
which is valid for $z, z' \not\in \bm z$. In view of the constraint $\mathcal C_{\bar z} \approx 0$ in \eqref{primary constraints}, this is equivalent to the non-ultralocal algebra \eqref{rs algebra intro} with $\R$-matrix as in \eqref{R matrix intro}.

The slightly unconventional factor of $2 \pi$ in \eqref{R matrix intro} matches with the conventions of \cite{Vicedo:2017cge}, where \eqref{rs algebra intro} was derived from purely algebraic considerations, as we shall recall in \S\ref{sec: NUL algebra}. Note that here $\delta_{\sigma\sigma'}$ denotes the Dirac comb, whereas in \cite{Vicedo:2017cge} we used it to denote the unnormalised Dirac comb, which in the present conventions is $2 \pi \delta_{\sigma\sigma'}$.

\medskip

We have now imposed the constraint $\gamma \approx 0$ strongly, or more precisely $\langle\!\langle \varepsilon, \gamma \rangle\!\rangle \approx 0$ for every $\varepsilon \in C^\infty(S^1 \times \CP, \g)_\omega$. Using the gauge fixing condition \eqref{holomorphic gauge} this gives
\begin{equation} \label{Pi zb mero}
\langle\!\langle \varepsilon, \partial_{\bar z} \Pi_{\bar z} \rangle\!\rangle \approx 0.
\end{equation}
It follows that $\Pi_{\bar z}$ is meromorphic on $\CP$ with the same pole structure as $\varphi$. By virtue of the definition of the constraint $\mathcal C_{\bar z} \approx 0$ in \eqref{primary constraints}, this is equivalent to \eqref{local condition}.

\begin{remark}
The condition \eqref{local condition} can also be seen in the Lagrangian formalism from the equation of motion $\omega \wedge F = 0$. In the gauge \eqref{holomorphic gauge} this implies that the connection $A_\tau d \tau + A_\sigma d \sigma$ is flat and that $\varphi \partial_{\bar z} A_\sigma = \varphi \partial_{\bar z} A_\tau = 0$. In other words, $A_\sigma$ and $A_\tau$ are meromorphic with poles at the zeroes of $\varphi$ (with the order of each pole coinciding with the multiplicity of the corresponding zero of $\varphi$).
\end{remark}

\subsection{Fixing the Lagrange multiplier} \label{sec: fix Lagr}

Note that we still have the first class primary
constraint $\Pi_\tau \approx 0$. The effect of the corresponding gauge transformation is just to change the Lagrange multiplier $A_\tau$ in the bulk Hamiltonian. We shall impose it strongly by fixing the Lagrange multiplier.

\medskip

Let $\bm \zeta$ denote the set of zeroes of $\omega$. We shall assume, for the sake of clarity of the presentation, that $\bm \zeta \subset \CC$, \emph{i.e.} infinity is not a zero, and moreover that all the zeroes are simple. The latter means that $\varphi(x) = 0$ while $\varphi'(x) \neq 0$ for $x \in \bm \zeta$. The arguments given below and in \S\ref{sec: reduced dynamics} generalise straightforwardly to the generic case.

Having imposed the constraint $\mathcal C_{\bar z} \approx 0$ in \eqref{primary constraints} strongly we have
\begin{equation} \label{Pi sigma explicit}
A_\sigma(\sigma, z) \approx \frac{4 \pi \ii}{\varphi(z)} \Pi_{\bar z}(\sigma, z) = \sum_{x \in \bm \zeta} \frac{4 \pi \ii}{\varphi'(x)} \frac{\Pi_{\bar z}(\sigma, x)}{z - x}.
\end{equation}
The second equality is obtained by performing a partial fraction expansion, noting that $\Pi_{\bar z}$ and $\varphi$ have the same pole structure. The explicit form above follows from assuming that $\varphi$ has simple zeroes at points in the set $\bm \zeta \subset \CC$. Note that \eqref{Pi sigma explicit} is exactly equation (2.39) from \cite{Delduc:2019bcl}, in view of the discussion of \S\ref{sec: NUL algebra} below.

Fix a set $\{ \epsilon_x \}_{x \in \bm \zeta}$ of complex numbers. We shall use the gauge fixing condition
\begin{equation} \label{gauge fix Pi tau}
A_\tau(\sigma, z) \approx - \sum_{x \in \bm \zeta} \frac{4 \pi \ii \epsilon_x}{\varphi'(x)} \frac{\Pi_{\bar z}(\sigma, x)}{z - x}.
\end{equation}
(Note that this coincides, up to a sign in the definition of the $\epsilon_x$, with (2.40) from \cite{Delduc:2019bcl} by the same remark as for \eqref{Pi sigma explicit} above.)
In other words, we take a linear combination of the singular parts at each $x \in \bm \zeta$ of the partial fraction decomposition \eqref{Pi sigma explicit} with coefficients $\epsilon_x$.
Since we are setting the Lagrange multiplier $A_\tau$ equal to a meromorphic function with poles in $\bm\zeta$, we are technically only specifying its `boundary value' at the points in $\bm\zeta$. In any case, there is no need to specify its value as a whole on $\CP$ since we are already working on the constraint surface $\gamma \approx 0$.
We will motivate the choice \eqref{gauge fix Pi tau} shortly in \S\ref{sec: reduced dynamics}, but for the time being it is interesting to compare with the choices made in \cite{Costello:2019tri}.

\medskip

To compare with \cite{Costello:2019tri}, let us split the set $\bm \zeta$ into two disjoint subsets as $\bm \zeta = \bm \zeta_+ \sqcup \bm \zeta_-$ and take $\epsilon_x = \pm 1$ for $x \in \bm \zeta_\pm$. Let us also note in passing that the latter condition was shown in \cite{Delduc:2019bcl} to imply that the resulting model is relativitistic. It follows from comparing \eqref{Pi sigma explicit} with \eqref{gauge fix Pi tau} that $A_\tau \pm A_\sigma$ is regular at each $x \in \bm \zeta_\pm$ and has a simple pole at every $x \in \bm \zeta_\mp$. This is to be compared with the boundary conditions imposed on the fields $A_\tau \pm A_\sigma$ at the zeroes of $\omega$ in \cite{Costello:2019tri}, where $|\bm\zeta|$ is even and $|\bm\zeta_+| = |\bm\zeta_-|$.
Note, however, that by contrast with \cite{Costello:2019tri} we do not choose to work in a gauge in which the pair of fields $A_\sigma \pm A_\tau$ both vanish at the poles of $\omega$. We will come back to this point in \S\ref{sec: formal vs realisations} below. Choosing the right gauge in the Hamiltonian formalism is essential since it is known, see \emph{e.g.} \cite{BBT-book}, that the form of the Poisson bracket \eqref{rs algebra intro} \--- \eqref{R matrix intro} is very sensitive to this choice.

Introducing a Dirac bracket to impose $\Pi_\tau \approx 0$ strongly, together with its gauge fixing condition \eqref{gauge fix Pi tau}, it is immediate that the Dirac bracket \eqref{PB of Pi} is unmodified.

\subsection{Reduced dynamics} \label{sec: reduced dynamics}

In a classical field theory with no local degrees of freedom, such as \eqref{holo CS action}, it is the choice of boundary condition on the Lagrange multipliers in the Hamiltonian, such as $A_\tau$ here, which completely determines the dynamics on the reduced phase space.
In this sense, the gauge fixing condition \eqref{gauge fix Pi tau} was chosen so as to produce the correct dynamics on the reduced phase space, as we now show.

\medskip

The variation of the bulk Hamiltonian \eqref{Hamiltonian v0} reads
\begin{align*}
\delta H_{\rm bulk} &= \langle\!\langle \gamma, \delta A_\tau \rangle\!\rangle
+ \langle\!\langle \mbox{\small $\frac{\ii}{2 \pi}$} \varphi \partial_\sigma A_\tau + 2 [A_\tau, \Pi_{\bar z}], \delta A_{\bar z} \rangle\!\rangle + 2 \langle\!\langle [A_{\bar z}, A_\tau] + \partial_{\bar z} A_\tau, \delta \Pi_{\bar z} \rangle\!\rangle.
\end{align*}
The first term vanishes on the constraint surface. Among all of the other terms, the only potentially problematic one is the one involving $\partial_{\bar z} A_\tau$ since it could correspond to a `boundary term', cf. Remark \ref{rem: boundary terms}. And indeed, by using the explicit form of the gauge fixing condition \eqref{gauge fix Pi tau} we can rewrite it as 
\begin{align*}
2 \langle\!\langle \partial_{\bar z} A_\tau, \delta \Pi_{\bar z} \rangle\!\rangle &= - 4 \pi \sum_{x \in \bm \zeta} \frac{4 \pi \epsilon_x}{\varphi'(x)} \int_{S^1 \times \CP} d\sigma \wedge dz \wedge d\bar z \, \delta_{z x} \langle \Pi_{\bar z}(\sigma, x), \delta \Pi_{\bar z}(\sigma, z) \rangle\\
&= \delta \Bigg( \!\! - \ha \sum_{x \in \bm \zeta} \frac{\epsilon_x}{\varphi'(x)} \int_{S^1} d\sigma \, \langle 4 \pi \Pi_{\bar z}(\sigma, x), 4 \pi \Pi_{\bar z}(\sigma, x) \rangle \Bigg).
\end{align*}
This suggests adding a boundary term to the bulk Hamiltonian $H_{\rm bulk}$, given in \eqref{Hamiltonian v0}, to cancel off this boundary term in the above variation $\delta H_{\rm bulk}$. Explicitly, we define the new Hamiltonian
\begin{equation*}
H \coloneqq \langle\!\langle A_\tau, \gamma \rangle\!\rangle + \ha \sum_{x \in \bm \zeta} \frac{\epsilon_x}{\varphi'(x)} \int_{S^1} d\sigma \langle 4\pi \Pi_{\bar z}(\sigma, x), 4\pi\Pi_{\bar z}(\sigma, x) \rangle,
\end{equation*}
which is now differentiable in the sense of \cite{Regge:1974zd}, see \S\ref{sec: diff func}.

The Hamiltonian on the reduced phase space is then given by
\begin{equation*}
H \approx \ha \sum_{x \in \bm \zeta} \frac{\epsilon_x}{\varphi'(x)} \int_{S^1} d\sigma \langle 4\pi \Pi_{\bar z}(\sigma, x), 4\pi\Pi_{\bar z}(\sigma, x) \rangle.
\end{equation*}
This can equally be rewritten as
\begin{equation} \label{Ham reduced PS}
H \approx \sum_{x \in \bm \zeta} \epsilon_x \res_x \bigg( \ha \varphi(z)^{-1} \int_{S^1} d\sigma \langle 4\pi \Pi_{\bar z}(\sigma, z), 4\pi\Pi_{\bar z}(\sigma, z) \rangle \bigg) dz.
\end{equation}
which is equivalent to \eqref{Ham intro} using the constraint $\mathcal C_{\bar z} \approx 0$. In this final form \eqref{Ham reduced PS}, it is straightforward to show that the result also holds, as written, in the more generic situation when $\omega$ is allowed to have multiple zeroes including at infinity.

\section{Connection with affine Gaudin models} \label{sec: affine Gaudin}

We showed in \S\ref{sec: gauge fixing} that the non-ultralocal Poisson algebra \eqref{rs algebra intro}, with $\R$-matrix given by \eqref{R matrix intro}, naturally arises as the Poisson structure on the reduced phase space of holomorphic Chern-Simons theory. We went on to show in \S\ref{sec: reduced dynamics} that for a suitable choice of gauge fixing conditions (closely related to conditions imposed in \cite{Costello:2019tri}), the Hamiltonian on the reduced phase space takes the very specific form \eqref{Ham intro}.

By contrast, in \cite{Vicedo:2017cge} we gave a very different, more algebraic, interpretation of this same non-ultralocal Poisson algebra \eqref{rs algebra intro} \--- \eqref{R matrix intro} and Hamiltonian \eqref{Ham intro}.
We will briefly review this below. In short, classical integrable field theories with properties \eqref{rs algebra intro} \--- \eqref{Ham intro} can equally be understood as particular representation of generalised (non-cyclotomic) affine Gaudin models.

\subsection{Non-ultralocal algebra} \label{sec: NUL algebra}

The object which naturally enters in the formalism of \cite{Vicedo:2017cge} is not so much the field $A_\sigma$ but rather the combination $\mathcal L \coloneqq 4 \pi \ii \Pi_{\bar z} = \varphi A_\sigma$. Its Poisson bracket, which follows immediately form \eqref{PB of Pi}, can be written as
\begin{align} \label{PB Lax matrix}
\{ \L_\1(\sigma, z), \L_\2(\sigma', z') \} &= \bigg[2 \pi \frac{C_{\1\2} \delta_{\sigma\sigma'}}{z'-z}, \varphi(z) \partial_\sigma + \L_\1(\sigma, z) \bigg] \notag\\
&\qquad\qquad\qquad + \bigg[2 \pi \frac{C_{\1\2} \delta_{\sigma\sigma'}}{z'-z}, \varphi(z') \partial_{\sigma'} + \L_\1(\sigma', z') \bigg].
\end{align}
We have explicitly removed the superscript `$\star$' on the Poisson bracket since in what follows we no longer want to think of it as a Dirac bracket on a reduced phase space.

To explain the origin of the Poisson bracket \eqref{PB Lax matrix} from Gaudin models associated with the untwisted affine Kac-Moody algebra $\tg \coloneqq \g \otimes \CC[t, t^{-1}] \oplus \CC \cent \oplus \CC \cocent$, we briefly recall how these are defined.

\medskip

Let $\{ I_a \}$ be a basis of $\g$ and denote $\{ I^a \}$ its dual basis with respect to the bilinear form $\langle \cdot, \cdot \rangle$. Note that, in terms of these, we can write the split Casimir of $\g$ introduced in \S\ref{sec: phase space} as $C = I_a \otimes I^a$, where the sum over the repeated index $a$ is implicit.

A basis $\{ I_{\widetilde a} \}$ of $\tg$ is then given by $I_{a, -n} \coloneqq I_a \otimes t^{-n}$ for $n \in \ZZ$ together with $\cent$ and $\cocent$. Its dual basis with respect to the standard bilinear form $(\cdot | \cdot) : \tg \times \tg \to \CC$ on $\tg$, which we denote by $\{ I^{\widetilde a} \}$, consists of $I^a_n \coloneqq I^a \otimes t^n$ for $n \in \ZZ$ together with $\cocent$ and $\cent$.

Now the Lax matrix of the affine Gaudin model takes the form
\begin{equation} \label{Gaudin Lax matrix}
L(z) \coloneqq I_{\widetilde a} \otimes \mathcal L^{\widetilde a}(z)
\end{equation}
where the infinite sum over the repeated index $\widetilde a$ is implicit. The $\mathcal L^{\widetilde a}(z)$ are given by very explicit rational functions on $\CP$ which are valued in the algebra of observables $\mathcal A$ of the Gaudin model. For instance, in the simplest case of Gaudin models with regular singularities the algebra of observables $\mathcal A$ is a completion of $S(\tg)^{\otimes N}$ where $N \in \ZZ_{\geq 1}$ is the number of sites, which are located at $\bm z = \{ z_i \}_{i=1}^N$. We then have
\begin{equation*}
\mathcal L^{\widetilde a}(z) = \sum_{i=1}^N \frac{I^{\widetilde a(i)}}{z - z_i}
\end{equation*}
where $I^{\widetilde a(i)}$ denotes the copy of the basis element $I^{\widetilde a} \in \tg$ in the $i^{\rm th}$ copy of $S(\tg)$ in the $N$-fold tensor product $S(\tg)^{\otimes N}$.
Explicit and simple expressions for $\mathcal L^{\widetilde a}(z)$ also exist for other generalisations of the Gaudin model. However, since these are not directly relevant for the present discussion, we refer the reader to \cite{Vicedo:2017cge} for the details. For our purposes, the key property that we shall need of these functions is that the Poisson brackets of the fundemantal fields of the Gaudin model (in the above example these are $I^{\widetilde a(i)}$) can be packaged into the following form \cite{Vicedo:2017cge}
\begin{equation} \label{Gaudin Lax PB}
\{ L_\1(z), L_\2(z') \} = \bigg[ \frac{\widetilde C_{\1\2}}{z'-z}, L_\1(z) \bigg] + \bigg[ \frac{\widetilde C_{\1\2}}{z'-z}, L_\2(z') \bigg],
\end{equation}
where $\widetilde C \coloneqq I_{\widetilde a} \otimes I^{\widetilde a}$ is the split Casimir of $\tg$.

\medskip

The connection with \eqref{PB Lax matrix} is now apparent. Explicitly, let us consider the natural representation $\varrho$ of $\tg$ in terms of $\g$-valued connections on $S^1$, given explicitly in the basis $\{ I_{\widetilde a} \}$ by
\begin{equation*}
I_{a, -n} \longmapsto I_a \otimes e^{- \ii n \sigma}, \qquad
\cent \longmapsto 0, \qquad
\cocent \longmapsto - \ii \partial_\sigma,
\end{equation*}
where $\sigma$ is a coordinate on $S^1 = \RR / 2 \pi \ZZ$. Applying $\varrho$ to both tensor factors of the split Casimir of $\tg$ yields
\begin{equation*}
\widetilde C = \cent \otimes \cocent + \cocent \otimes \cent + \sum_{n \in \ZZ} I_{a, -n} \otimes I^a_n \; \xmapsto{\varrho \otimes \varrho} \; (I_a \otimes I^a) \sum_{n \in \ZZ} e^{- \ii n(\sigma - \sigma')} = 2 \pi C \delta_{\sigma \sigma'}.
\end{equation*}
As recalled above, the definition of $\delta_{\sigma\sigma'}$ used here is $\mbox{\small $\frac{1}{2\pi}$}$ times the one in \cite{Vicedo:2017cge}.

Likewise, applying $\varrho$ to the first tensor factor of the formal Lax operator \eqref{Gaudin Lax matrix} gives
\begin{align*}
L(z) &= \cent \otimes \mathcal D(z) + \cocent \otimes \mathcal K(z) + \sum_{n \in \ZZ} I_{a, -n} \otimes \mathcal L^a_n(z)\\
&\qquad\quad \xmapsto{\varrho \otimes \id} \; - \ii \partial_\sigma \otimes \mathcal K(z) + I_a \sum_{n \in \ZZ} e^{- \ii n \sigma} \otimes \mathcal L^a_n(z),
\end{align*}
where again we refer to \cite{Vicedo:2017cge} for the explicit forms of the rational functions $\mathcal D(z)$, $\mathcal K(z)$ and $\mathcal L^a_n(z)$ valued in the algebra of observables $\mathcal A$ of the affine Gaudin model.

\medskip

To describe a specific classical integrable field theory we should also introduce a representation $\hat \pi$ of the Poisson algebra $\mathcal A$. This should send $\mathcal K(z)$, which is valued in the centre of $\mathcal A$, to a complex number valued rational function. In the example of a Gaudin model with regular singularities mentioned above this takes the form
\begin{equation*}
\mathcal K(z) = \sum_{i=1}^N \frac{\cent^{(i)}}{z - z_i},
\end{equation*}
and the central elements $\cent^{(i)}$ should be realised as complex numbers.
Furthermore, $\hat\pi$ should realise each $\mathcal L^a_n(z)$, $n \in \ZZ$ in terms of the Fourier modes of the various fields of the classical integrable field theory in question. Explicitly, $\hat \pi$ is given by
\begin{equation} \label{rep pi}
\mathcal K(z) \longmapsto \ii \varphi(z), \qquad
I_a \sum_{n \in \ZZ} e^{- \ii n \sigma} \otimes \mathcal L^a_n(z) \longmapsto \mathcal L(\sigma, z),
\end{equation}
where $\mathcal L(z)$ is the $\g$-valued Lax matrix of the classical integrable field theory and $\varphi(z)$ is its twist function. Combining this with the representation $\varrho$ we have
\begin{equation*}
L(z) \; \xmapsto{\varrho \otimes \hat \pi} \; \varphi(z) \partial_\sigma + \L(\sigma, z).
\end{equation*}
In other words, the twist function naturally arises as one of the components of the Lax matrix of the affine Gaudin model.

Applying $\varrho$ to the first and second tensor factors of the Poisson bracket relation \eqref{Gaudin Lax PB}, labelled respectively by $\1$ and $\2$, as well as applying $\hat \pi$ to the third factor which is not explicitly labelled, we now obtain the non-ultralocal Poisson algebra \eqref{PB Lax matrix}.

\subsection{Quadratic Hamiltonians}

So far we have only described, though somewhat implicitly (but more explicitly in the case of regular singularities), the kinematics of affine Gaudin models.

The dynamics of an affine Gaudin model is defined by its \emph{quadratic Hamiltonians}. These are conveniently defined, by using the Lax matrix \eqref{Gaudin Lax matrix}, as the coefficients in the partial fraction expansion of the rational function
\begin{equation} \label{S1 def}
S_1(z) \coloneqq \ha (L(z) | L(z)) = \mathcal K(z) \mathcal D(z) + \ha \sum_{n \in \ZZ} \langle I_a, I_b \rangle \mathcal L^a_{-n}(z) \mathcal L^b_n(z),
\end{equation}
where the bilinear form $(\cdot | \cdot)$ on $\tg$ is being applied to the pair of first factors of the Lax matrices in \eqref{Gaudin Lax matrix}. It follows directly from \eqref{Gaudin Lax PB} that the quadratic Hamiltonians generate an abelian subalgebra of $\mathcal A$ \cite{Vicedo:2017cge}.

\medskip

Since \eqref{S1 def} is a rational function valued in $\mathcal A$, we can apply to it the representation $\hat\pi$ which, after also multiplying through by the inverse twist function, gives \cite{Delduc:2019bcl}
\begin{equation*}
\varphi(z)^{-1} \hat \pi \big( S_1(z) \big) = \hat \pi \big( \mathcal D(z) \big) + \ha \varphi(z)^{-1} \int_{S^1} d\sigma \langle \mathcal L(\sigma, z), \mathcal L(\sigma, z) \rangle.
\end{equation*}
The first term on the right hand side has poles only at the sites $\bm z = \{ z_i \}_{i =1}^N$, namely at the poles of the twist function $\varphi$, and is thus regular at the set $\bm \zeta$ of zeroes of $\varphi$. Taking the residue at any $x \in \bm \zeta$ we obtain
\begin{equation*}
\res_x \varphi(z)^{-1} \hat \pi \big( S_1(z) \big) dz = \res_x \bigg( \ha \varphi(z)^{-1} \int_{S^1} d\sigma \langle \mathcal L(\sigma, z), \mathcal L(\sigma, z) \rangle \bigg) dz.
\end{equation*}
Recalling that we have already identified $\mathcal L = 4 \pi \ii \Pi_{\bar z}$ in \S\ref{sec: NUL algebra}, it now follows that the Hamiltonian \eqref{Ham reduced PS} of holomorphic Chern-Simons theory on the reduced phase space is given by a linear combination of the quadratic Gaudin Hamiltonians, explicitly
\begin{equation*}
H \approx - \sum_{x \in \bm \zeta} \epsilon_x \res_x \varphi(z)^{-1} \hat \pi \big( S_1(z) \big) dz.
\end{equation*}
This completes the proof of the main result, namely that the classical integrable field theory on the reduced phase space of holomorphic Chern-Simons theory identified in \S\ref{sec: Ham CS theory} can indeed be described as a realisation of an affine Gaudin model.

\medskip

Let us note in passing that higher-spin local integrals of motion in classical Gaudin models of affine type are also intrinsically associated with the set $\bm\zeta$. Indeed, explicit expressions for these were constructed in \cite{Lacroix:2017isl}, in the case when $\g$ is of classical type, generalising the original construction of \cite{Evans:1999mj} on the principal chiral model. Specifically, there exists certain polynomials in the Lax matrix $\mathcal L(\sigma, z)$, whose degrees are related to the set of exponents of $\tg$, and the evaluation of which at the points in $\bm\zeta$ yield the higher local conserved charges.
It would be interesting to understand the appearance of these from the point of view of holomorphic Chern-Simons theory.

\section{Discussion} \label{sec: discussion}

\subsection{Formal Gaudin model and realisations} \label{sec: formal vs realisations}

Loosely speaking, one talks about a given classical integrable field theory as `being' an affine Gaudin model if it has all the properties listed in \eqref{rs algebra intro} \--- \eqref{Ham intro}.
However, it is convenient to distinguish the affine Gaudin model formulated at the abstract level of affine Kac-Moody algebras from the classical integrable field theory itself. For this reasons, quantities expressed at the level of Kac-Moody algebras were referred to as being \emph{formal} in \cite{Vicedo:2017cge}.

\medskip

As recalled in \S\ref{sec: affine Gaudin}, in order to go from the formal affine Gaudin model to a concrete classical integrable field theory, one needs to make a choice of representation $\hat \pi$ of the algebra of formal observables $\mathcal A$, cf. \eqref{rep pi}. And although the twist function $\varphi$ is an important ingredient in the definition of $\hat \pi$ it does not, by itself, define $\hat \pi$. Indeed, one also needs a \emph{realisation} of the formal fields of the Gaudin model in terms of the fundamental fields of a given theory, represented by the second equation in \eqref{rep pi}.

In particular, different classical integrable field theories may share the same twist function. Indeed, given a twist function with at most double poles, there are often various natural ways of defining a corresponding realisation $\hat\pi$. A list of possibilities, which is by no means complete, was given in \cite{Delduc:2019bcl}.

\medskip

One way of defining $\hat\pi$ is to try to associate with every double pole of $\varphi$, or with pairs of simple poles of $\varphi$, a copy of the cotangent bundle $T^\ast \mathcal L G$ of the loop group $\mathcal L G$ where $G$ is a real Lie group with Lie algebra $\g$, which we take here to be real. A general recipe for doing so was given in \cite{Vicedo:2015pna} building on the earlier constructions in \cite{Delduc:2013fga, Delduc:2014kha}.

Concretely, the group valued field $g_i$ parameterising the base of the copy of $T^\ast \mathcal L G$ associated with a given double pole $z_i$ of $\varphi$ can be defined by the requirement that the gauge transformation of the Lax matrix by $g_i$ vanishes at $z_i$. Likewise, the group valued field $g_i$ associated with simple poles $z_i^\pm$ %and $z_i^-$
is defined by requiring that the gauge transformation of the Lax matrix by $g_i$ evaluated at the pair of points $z^\pm_i$ takes value in a subalgebra complemenetary to $\g$ in $\g^\CC$ or to $\g_{\rm diag}$ in $\g \oplus \g$. See \cite{Vicedo:2015pna} for details.

The proposals of \cite{Costello:2019tri} for constructing the group valued fields (called $\sigma_i$ there) in both the double and simple pole cases, \emph{i.e.} rational and trigonometric cases, are very reminiscent of the above general constructions. This serves to highlight again the very close similarity between the two formalisms of \cite{Costello:2019tri} and \cite{Vicedo:2017cge}.

\medskip

Let us also mention that one particular family of classical integrable field theories that were shown in \cite{Delduc:2019bcl} (see also \cite{Vicedo:2015pna}) to be realisations of affine Gaudin models are the so called `$\lambda$-deformations' of the principal chiral model \cite{Sfetsos:2013wia}, of the symmetric space $\sigma$-models \cite{Hollowood:2014rla} and also of the semi-symmetric space $\sigma$-models \cite{Hollowood:2014qma}; see also \cite{Benitez:2019oaw} for the $\lambda$-deformation of the pure-spinor superstring on $AdS_5 \times S^5$. It was argued in \cite{Schmidtt:2017ngw,Schmidtt:2018hop} that the $\lambda$-deformation can be seen as the theory at the boundary of a `doubled' ordinary Chern-Simons theory. It would be interesting to understand the connection with the present analysis in the context of $\lambda$-deformations.

\subsection{Dihedral equivariance} \label{sec: dihedral}

It will be interesting to generalise the analysis of the present note to holomorphic Chern-Simons theory on the orbifold $\Sigma \times \CP/\ZZ_T$ for $T \in \ZZ_{\geq 2}$, where $\ZZ_T$ here only acts on $\CP$ by contrast with the orbifolds considered in \cite{Bittleston:2019gkq}. This should amount to $A_\sigma$ being equivariant under an action of the cyclic group $\ZZ_T$ in the sense that
\begin{equation} \label{cyclotomic}
\check\sigma(A_\sigma(\sigma, z)) = A_\sigma(\sigma, \omega z),
\end{equation}
where $\omega$ is a $T^{\rm th}$-root of unity and $\check\sigma$ a $\ZZ_T$-automorphism of $\g$.

In the language of \cite{Vicedo:2017cge}, this would then correspond to considering the family of $\ZZ_T$-cyclotomic affine Gaudin models. The latter encompases all symmetric and semi-symmetric space $\sigma$-models and in particular the $\sigma$-model of the superstring on $AdS_5 \times S^5$ \cite{Vicedo:2010qd}, but also affine Toda field theories.

\medskip

We have also not addressed the issue of reality conditions here. In the setting of \cite{Vicedo:2017cge} these are characterised by the Lax matrix $A_\sigma$ also being equivariant under an action of $\ZZ_2$, namely
\begin{equation} \label{reality}
\check\tau(A_\sigma(\sigma, z)) = A_\sigma(\sigma, \bar z),
\end{equation}
where $\check\tau$ is an anti-linear involution of the complex Lie algebra $\g$, that specifies the choice of real form of $\g$.

The conditions \eqref{cyclotomic} and \eqref{reality} put together imply that $A_\sigma$ is, in fact, equivariant under an action of the dihedral group $D_{2T} = \ZZ_T \rtimes \ZZ_2$. It was shown, more precisely, in \cite{Vicedo:2017cge} that many classical integrable field theories of interest admit a description as \emph{dihedral} affine Gaudin models. It will be interesting to connect in detail such affine Gaudin models to holomorphic Chern-Simons theory. We leave this for future work.

\subsection{(Dis)order defects and (non-)ultralocality} \label{sec: (dis)order}

There are two types of classical integrable field theories discussed in \cite{Costello:2019tri}, corresponding to two types of surface defects, namely the \emph{order} and \emph{disorder} ones, that can be added to the holomorphic Chern-Simons theory described by the bulk action \eqref{holo CS action}. It follows from the results of this note that this dichotomy is essentially the same as the usual one between the \emph{ultralocal} and \emph{non-ultralocal} models.

\medskip

Indeed, the \emph{order defects} were considered only in the cases when $\omega$ has no zeroes, as in the original papers \cite{Costello:2013zra, Costello:2013sla, Witten:2016spx, Costello:2017dso,Costello:2018gyb} on lattice models \--- see \S\ref{sec: zeroes of omega} below. Among those, the two cases covered by the formalism of this note are $\omega = dz$ (rational) and $\omega = dz / z$ (trigonometric). In the former case $\varphi(z) = 1$ so that the $\delta'$-term in the Poisson bracket \eqref{rs algebra intro} of the Lax matrix is absent. In the latter case, the coefficient of the $\delta'$-term in \eqref{rs algebra intro} is constant, \emph{i.e.} independent of the spectral parameters, and can typically be eliminated by a suitable gauge transformation. A prime example of this, albeit in the cyclotomic case, is given by KdV theory \cite{Bazhanov:1994ft}.

\medskip

As noted in \cite{Costello:2019tri}, however, the collection of classical integrable field theories that can be described using order defects is very limited. Indeed, most theories of interest are described instead, in the language of \cite{Costello:2019tri}, using so called \emph{disorder defects}. These were considered in the case when the 1-form $\omega$ has zeroes. As we have shown, this is in perfect agreement with the observation made in \cite{Vicedo:2017cge} that a very large family of classical integrable field theories are described by affine Gaudin models, which are intrinsically non-ultralocal. Indeed, the fact that many known non-ultralocal models were recovered in \cite{Costello:2019tri}, including the multi-parameter family of coupled integrable $\sigma$-models introduced in \cite{Delduc:2018hty}, is what originally prompted us to seek a deeper connection between the formalisms of \cite{Costello:2019tri} and \cite{Vicedo:2017cge}.

\medskip

Turning to the problem of quantising these classical integrable field theories, one can expect the quantum inverse scattering method \cite{QISM1,QISM2,QISM3,QISM4,QISM5}, \emph{i.e.} RTT formalism, to apply as usual in the ultralocal setting. In particular, this formalism should have a reinterpretation in the language of holomorphic Chern-Simons theory as was the case for lattice models in \cite{Costello:2018gyb}.

In the non-ultralocal setting, however, we expect new techniques to be required, which are ultimately related to the problem of $\omega$ having zeroes.

\subsection{Zeroes of the differential $\omega$} \label{sec: zeroes of omega}

The presence of zeroes in $\omega$ is known to pose problems in the perturbative quantisation of holomorphic Chern-Simons theory. Indeed, it was argued heuristically, \emph{e.g.} in \cite{Costello:2017dso}, that since the action \eqref{holo CS action} depends on $\omega$ through the ratio $\omega/\hbar$, its zeroes correspond to points where $\hbar \to \infty$. In light of the discussion of \S\ref{sec: (dis)order}, this issue can be seen as a reformulation of the long-standing open problem of quantising non-ultralocal integrable field theories, which in turn is equivalent to the problem of quantising (dihedral) affine Gaudin models \cite{Vicedo:2017cge}.

It is interesting that in the lattice model context of \cite{Costello:2013zra, Costello:2013sla, Witten:2016spx, Costello:2017dso,Costello:2018gyb}, restricting attention to Riemann surfaces $C$ admitting a non-vanishing differential $\omega$, so as to avoid these difficulties, has led to rediscovering the classification of skew-symmetric solutions to the Yang-Baxter equation due to Belavin and Drinfel'd \cite{Belavin-Drinfeld}.

\medskip

By contrast, the presence of zeroes in $\omega$ is clearly needed in the context of classical integral field theories. In the language of \cite{Costello:2019tri}, it is thus expected that quantising non-ultralocal integrable field theories will require a non-perturbative definition of quantum holomorphic Chern-Simons theory with action \eqref{holo CS action} for generic $\omega$.

On the other hand, approaching the problem from the perspective of affine Gaudin models, we anticipate from \cite{Lacroix:2018fhf, Lacroix:2018itd} that in studying \emph{quantum} Gaudin models associated with the affine Kac-Moody algebra $\tg$, the role of the zeroes of the twist function $\varphi$ should be replaced by twisted homology cycles in $\CP \setminus \bm z$. This may also shed some light on how to tackle the problem from the point of view of holomorphic Chern-Simons theory.

\medskip

Finally, it is also expected that Langlands duality should play a central role in the study of Gaudin models in affine type, see for instance \cite{FFsolitons, FH, Lacroix:2018fhf, Lacroix:2018itd}, by direct analogy with the well-studied case of Gaudin models in finite type \cite{FFR, Fre1, Fre2, MukhinVarchenko1, MukhinVarchenko2}. It would therefore be very interesting to see the emergence of Langlands duality also from the point of view of holomorphic Chern-Simons theory.


\begin{thebibliography}{99}

\bibitem[BBT]{BBT-book}
  O.~Babelon, D.~Bernard and M.~Talon,
  \emph{Introduction to classical integrable systems},
  Cambridge University Press (2003).

\bibitem[B1]{Banados:1994tn}
  M.~Ba\~nados,
  \emph{Global charges in Chern-Simons field theory and the (2+1) black hole},
  Phys.\ Rev.\ D {\bf 52} (1996) 5816.
    
\bibitem[B2]{Banados:1998gg}
  M.~Ba\~nados,
  \emph{Three-dimensional quantum geometry and black holes},
  AIP Conf.\ Proc.\  {\bf 484}, no.1 (1999) 147.

\bibitem[BR]{Banados:2016zim}
  M.~Ba\~nados and I.~A.~Reyes,
  \emph{A short review on Noether’s theorems, gauge symmetries and boundary terms},
  Int.\ J.\ Mod.\ Phys.\ D {\bf 25} (2016) no.10, 1630021.

\bibitem[BLZ]{Bazhanov:1994ft}
  V.~V.~Bazhanov, S.~L.~Lukyanov and A.~B.~Zamolodchikov,
  \emph{Integrable structure of conformal field theory, quantum KdV theory and thermodynamic Bethe ansatz},
  Commun.\ Math.\ Phys.\  {\bf 177} (1996) 381.

\bibitem[BD]{Belavin-Drinfeld}
  A.~A.~Belavin and V.~G.~Drinfeld,
  \emph{Triangle equations and simple Lie algebras}, Classic Reviews in Mathematics and Mathematical Physics. 1. Amsterdam: Harwood Academic Publishers. vii, 91 p. (1998).

\bibitem[BeS]{Benitez:2019oaw}
  H.~A.~Ben\'{\i}tez and D.~M.~Schmidtt,
  \emph{$\lambda$-Deformation of the $AdS_{5}\times S^{5}$ Pure Spinor Superstring},
  arXiv:1907.13197 [hep-th].

\bibitem[BS]{Bittleston:2019gkq}
  R.~Bittleston and D.~Skinner,
  \emph{Gauge Theory and Boundary Integrability},
  JHEP {\bf 1905} (2019) 195.

\bibitem[BH1]{Brown:1986nw}
  J.~D.~Brown and M.~Henneaux,
  \emph{Central Charges in the Canonical Realization of Asymptotic Symmetries: An Example from Three-Dimensional Gravity},
  Commun.\ Math.\ Phys.\  {\bf 104} (1986) 207.

\bibitem[BH2]{Brown:1986ed}
  J.~D.~Brown and M.~Henneaux,
  \emph{On the Poisson Brackets of Differentiable Generators in Classical Field Theory},
  J.\ Math.\ Phys.\  {\bf 27} (1986) 489.

\bibitem[C1]{Costello:2013zra}
  K.~Costello,
  \emph{Supersymmetric gauge theory and the Yangian},
  arXiv:1303.2632 [hep-th].

\bibitem[C2]{Costello:2013sla}
  K.~Costello,
  \emph{Integrable lattice models from four-dimensional field theories},
  Proc.\ Symp.\ Pure Math.\  {\bf 88} (2014) 3.

\bibitem[CWY1]{Costello:2017dso}
  K.~Costello, E.~Witten and M.~Yamazaki,
  \emph{Gauge Theory and Integrability, I},
  ICCM Not. {\bf 6} (2018) 46--191.

\bibitem[CWY2]{Costello:2018gyb}
  K.~Costello, E.~Witten and M.~Yamazaki,
  \emph{Gauge Theory and Integrability, II},
  ICCM Not. {\bf 6} (2018) 120--149.

\bibitem[CY]{Costello:2019tri}
  K.~Costello and M.~Yamazaki,
  \emph{Gauge Theory And Integrability, III},
  arXiv:1908.02289 [hep-th].

\bibitem[DLMV1]{Delduc:2018hty}
  F.~Delduc, S.~Lacroix, M.~Magro and B.~Vicedo,
  \emph{Integrable Coupled $\sigma$ Models},
  Phys.\ Rev.\ Lett.\  {\bf 122} (2019) no.4,  041601.

\bibitem[DLMV2]{Delduc:2019bcl}
  F.~Delduc, S.~Lacroix, M.~Magro and B.~Vicedo,
  \emph{Assembling integrable $\sigma$-models as affine Gaudin models},
  JHEP {\bf 1906} (2019) 017.

\bibitem[DMV1]{Delduc:2013fga}
  F.~Delduc, M.~Magro and B.~Vicedo,
  \emph{On classical $q$-deformations of integrable sigma-models},
  JHEP {\bf 1311} (2013) 192.

\bibitem[DMV2]{Delduc:2014kha}
  F.~Delduc, M.~Magro and B.~Vicedo,
  \emph{Derivation of the action and symmetries of the $q$-deformed $AdS_{5} \times S^{5}$ superstring},
  JHEP {\bf 1410} (2014) 132.

\bibitem[D]{QISM4}
  V. G. Drinfeld,
  \emph{Quantum Groups},
  J. Sov. Math. {\bf 41} (1988) 898.

\bibitem[EHMM]{Evans:1999mj}
  J.~M.~Evans, M.~Hassan, N.~J.~MacKay and A.~J.~Mountain,
  \emph{Local conserved charges in principal chiral models},
  Nucl.\ Phys.\ B {\bf 561} (1999) 385.

\bibitem[FST]{QISM3}
  L.~Faddeev, E.~K.~Sklyanin and L.~Takhtajan,
  \emph{The Quantum Inverse Problem Method. 1},
  Theor. Math. Phys. {\bf 40} (1980) 688.

\bibitem[FRT]{QISM5}
  L.~Faddeev, N.~Reshitikhin and L.~Takhtajan,
  \emph{Quantization of Lie Groups and Lie Algebras},
  Algebr. Analiz. {\bf 1}, LOMI-E-87-14 (1987).

\bibitem[FT]{QISM1}
  L.~Faddeev and L.~Takhtajan,
  \emph{The quantum method of the inverse problem and the Heisenberg XYZ-model},
  Russ. Math. Surveys {\bf 34:5} (1979) 1168.

\bibitem[FF]{FFsolitons}
  B.~Feigin and E.~Frenkel,
  \emph{Quantization of soliton systems and Langlands duality}, Adv. Stud. Pure. Math. {\bf 61}, Math. Soc. Japan, Tokyo, 2011.

\bibitem[FFR]{FFR}
  B.~Feigin, E.~Frenkel and N.~Reshetikhin,
  \emph{Gaudin model, Bethe ansatz and correlation functions at the critical level},
  Commun. Math. Phys. {\bf 166} (1994), 27--62.

\bibitem[F1]{Fre1}
  E.~Frenkel,
  \emph{Opers on the projective line, flag manifolds and Bethe ansatz},
  Mosc. Math. J. {\bf 4} (2004), no. 3, 655--705, 783.

\bibitem[F2]{Fre2}
  E.~Frenkel,
  \emph{Gaudin model and opers},
  Infinite dimensional algebras and quantum integrable systems, Progr. Math., vol. {\bf 237}, Birkhäuser, Basel, 2005, pp. 1--58.

\bibitem[FH]{FH}
  E.~Frenkel and D.~Hernandez,
  \emph{Spectra of quantum KdV Hamiltonians, Langlands duality, and affine opers},
  Commun. Math. Phys. {\bf 362} (2018), no. 2, 362--361.

\bibitem[HMS1]{Hollowood:2014rla}
  T.~J.~Hollowood, J.~L.~Miramontes and D.~M.~Schmidtt,
  \emph{Integrable Deformations of Strings on Symmetric Spaces},
  JHEP {\bf 1411} (2014) 009.

\bibitem[HMS2]{Hollowood:2014qma}
  T.~J.~Hollowood, J.~L.~Miramontes and D.~M.~Schmidtt,
  \emph{An Integrable Deformation of the $AdS_5 \times S^5$ Superstring},
  J.\ Phys.\ A {\bf 47} (2014) no.49,  495402.
  
\bibitem[KS]{QISM2}
  P.~P.~Kulish and E.~K.~Sklyanin,
  \emph{Quantum inverse scattering method and the Heisenberg ferromagnet},
  Phys. Lett. {\bf A70} (1979) 461.

\bibitem[L]{Lacroix:2019xeh}
  S.~Lacroix,
  \emph{Constrained affine Gaudin models and diagonal Yang-Baxter deformations},
  arXiv:1907.04836 [hep-th].

\bibitem[LMV]{Lacroix:2017isl}
  S.~Lacroix, M.~Magro and B.~Vicedo,
  \emph{Local charges in involution and hierarchies in integrable sigma-models},
  JHEP {\bf 1709} (2017) 117.

\bibitem[LVY1]{Lacroix:2018fhf}
  S.~Lacroix, B.~Vicedo and C.~Young,
  \emph{Affine Gaudin models and hypergeometric functions on affine opers},
  Adv.\ Math.\  {\bf 350} (2019) 486.

\bibitem[LVY2]{Lacroix:2018itd}
  S.~Lacroix, B.~Vicedo and C.~A.~S.~Young,
  \emph{Cubic hypergeometric integrals of motion in affine Gaudin models},
  to appear in Adv. Theor. Math. Phys., arXiv:1804.06751 [math.QA].

\bibitem[M1]{Maillet1}
  J.~M.~Maillet,
  \emph{Kac-Moody algebra and extended Yang-Baxter relations in the O(N) non-linear sigma model},
  Phys. Lett. B {\bf 162} (1985) 137.

\bibitem[M2]{Maillet2}
  J.~M.~Maillet,
  \emph{New integrable canonical structures in two-dimensional models},
  Nucl. Phys. B {\bf 269} (1986) 54.
x
\bibitem[MV1]{MukhinVarchenko1}
  E.~Mukhin and A.~Varchenko,
  \emph{Critical points of master functions and flag varieties},
  Communication in Contempory Mathematics {\bf 6} (2004), no. 1, 111--163.

\bibitem[MV2]{MukhinVarchenko2}
  E.~Mukhin and A.~Varchenko,
  \emph{Miura Opers and Critical Points of Master Functions},
  Cent. Eur. J. Math. {\bf 3} (2005), 155--182.

\bibitem[RT]{Regge:1974zd}
  T.~Regge and C.~Teitelboim,
  \emph{Role of Surface Integrals in the Hamiltonian Formulation of General Relativity},
  Annals Phys.\  {\bf 88} (1974) 286.

\bibitem[S1]{Schmidtt:2017ngw}
  D.~M.~Schmidtt,
  \emph{Integrable Lambda Models And Chern-Simons Theories},
  JHEP {\bf 1705} (2017) 012.

\bibitem[S2]{Schmidtt:2018hop}
  D.~M.~Schmidtt,
  \emph{Lambda Models From Chern-Simons Theories},
  JHEP {\bf 1811} (2018) 111.

\bibitem[Sf]{Sfetsos:2013wia}
  K.~Sfetsos,
  \emph{Integrable interpolations: From exact CFTs to non-Abelian T-duals},
  Nucl.\ Phys.\ B {\bf 880} (2014) 225.

\bibitem[V1]{Vicedo:2010qd}
  B.~Vicedo,
  \emph{The classical R-matrix of AdS/CFT and its Lie dialgebra structure},
  Lett.\ Math.\ Phys.\  {\bf 95} (2011) 249.

\bibitem[V2]{Vicedo:2015pna}
  B.~Vicedo,
  \emph{Deformed integrable $\sigma$-models, classical $R$-matrices and classical exchange algebra on Drinfel’d doubles},
  J.\ Phys.\ A {\bf 48} (2015) no.35,  355203.

\bibitem[V3]{Vicedo:2017cge}
  B.~Vicedo,
  \emph{On integrable field theories as dihedral affine Gaudin models},
  Int. Math. Res. Not. {\bfseries rny128} (2018).

\bibitem[W]{Witten:2016spx}
  E.~Witten,
  \emph{Integrable Lattice Models From Gauge Theory},
  Adv.\ Theor.\ Math.\ Phys.\  {\bf 21} (2017) 1819.


\end{thebibliography}
\end{document}